\pdfoutput=1
\documentclass[%
 reprint,
nofootinbib,
 amsmath,amssymb,
 aps,
]{revtex4-1}

\usepackage[dvipdfmx]{graphicx}
\usepackage{dcolumn}
\usepackage{bm}
\usepackage{comment}
\usepackage{booktabs}
\usepackage{color}

\usepackage{braket}

\usepackage{color}

\begin{document}

\preprint{APS/123-QED}

\title{Decay properties of $N(1535)$ in the holographic QCD}

\author{Akihiro Iwanaka}
 \email{iwanaka@rcnp.osaka-u.ac.jp}
 \author{Daisuke Fujii}
 \email{daisuke@rcnp.osaka-u.ac.jp}
\author{Atsushi Hosaka}%
 \altaffiliation[Also at ]{Advanced Science Research Center, Japan Atomic Energy Agency, Tokai, Ibaraki 319-1195 Japan.}
  \email{hosaka@rcnp.osaka-u.ac.jp}
\affiliation{%
 Research Center for Nuclear Physics(RCNP), Osaka University, Ibaraki 567-0048, Japan.\\
}%

\date{\today}

\begin{abstract}
We study one pion emission decay of the first excited state of the nucleon with negative parity 
$N(1535) \equiv N^*$ in the holographic model of QCD.  
The excited state is described as a vibrational mode along the extra $z$ direction  in the five-dimensional space-time of the model.
We have obtained an analytic formula for the axial coupling of $g_A^{NN^*}$.
The off-diagonal axial coupling is obtained at the decaying pion momentum $|\bm{k}| = 448$ MeV as $g_A^{NN^*} \sim 0.32$, and hence a partial decay width $\Gamma_{N^* \to \pi N} \sim 30$ MeV, which is smaller than but reasonably compared to the experimental data.  
\end{abstract}

\maketitle

\section{\label{Introduction}Introduction}
Baryons at low energies show various features due to the rich structure of QCD.  
Ground state properties such as masses and magnetic moments are now 
computed directly from QCD by the first principle method of the lattice QCD.  
Yet, effective model approaches are useful to describe dynamical processes 
such as resonance formations or decays.  
A widely employed is the quark model, where resonances emerge as 
single particle excitations of quarks~\cite{Isgur:1978, Isgur:1979p, Isgur:1979g, Anthony:1983}.

When considering dynamical processes it is important to include 
interactions with pions because many resonances are formed and decay 
through pions.  
The importance of the pion dynamics is also expected by 
spontaneous breaking of chiral symmetry~\cite{Nambu:1961fr, Nambu:1961tp}.  
One way to include pions in baryon dynamics is realized by the Skyrme model, 
where baryons emerge as solitons of the non-linear sigma model~\cite{Skyrme:1962vh, Zahed:1986qz}.  
The approach has been justified in the large $N_c$ limit, where $N_c$ is the number
 of colors, and various 
baryon resonances appear as collective excitations~\cite{Witten:1979kh, Witten:1983tw, Witten:1983tx}. 
For instance, the $\Delta(1232)$ is a rotational excitation of the deformed 
hedgehog in the spin-isospin space, and 
the Roper resonance $N(1440)$ a monopole vibration of the radial motion~\cite{Kaulfuss:1985na}.  
Moreover, negative parity states may be
described as composite states of a ground state baryon 
and a negative parity meson such as $K\Sigma$ for $N(1535)$ and $\bar{K}N$ for $\Lambda(1450)$~\cite{Callan:1985hy, Callan:1987xt, Ezoe:2020piq}.  

A similar, but an alternative approach is the holographic QCD.  
A well-known is the Sakai-Sugimoto model, which has been used 
to investigate various non-perturbative properties of QCD ~\cite{Sakai:2004cn, Sakai:2005yt}. 
The model is based on flavor gauge theory in the five-dimensional space, 
the four-dimensional added space-time with one extra dimension ($z$-direction).  
A solitonic picture for baryons is then realized by an instanton 
in that five-dimensional space ~\cite{Hata:2007mb}. 
As in the Skyrme model baryon resonances are described by collective excitations of 
the hedgehog (Belavin--Polyakov--Schwartz--Tyupkin (BPST)) instanton.
A somewhat differently looking feature of this approach is that negative 
parity states are described by the collective vibration along the $z$-direction.  
Static properties have been investigated, with reasonable 
agreement with experimental data ~\cite{Hata:2007mb}. 

Now concerning dynamical properties, the one-pion decay of $N(1440)$ 
was investigated in Ref ~\cite{Fujii:2021}. 
A crucial point of this approach is that a finite value of the partial decay width 
is obtained in the long wave-length limit (of the pion) with good agreement with data. 
Moreover, a model independent relation was found for the  axial coupling constants between the nucleon 
and Roper resonance.  
These features sharply contrast with non-relativistic quark model calculations, 
where the decay is forbidden
\footnote{Higher order terms in the relativistic expansion can make it finite ~\cite{Arifi:2021orx}.}.

Now after having observed the good features in the decay of the $N(1440)$, 
in this paper we investigate the decay of the negative parity baryon $N(1535)$, 
which is the first negative parity resonance, in the holographic QCD, the state is described as 
a collective vibration along the $z$-direction.  
The method of the computation parallels the previous one ~\cite{Fujii:2021}.  
What are new here are the computation of the zeroth (time) component of the axial current,   
and the evaluation of the matrix elements in the $z$ valuable. 

This paper is organized as follows. 
In Section 2, we present the action used in this paper and obtain the classical 
instanton solution to the equations of motion. 
We then define the chiral currents and give their concrete expressions. 
In Section 3, we compute the axial coupling and decay width of $N(1535)$ in comparison with 
experimental data.  
The final section is for discussions and summary.  

\section{AXIAL CURRENT}
\subsection{Classical solutions and collective quantization}
We briefly review how baryons are formed in the Sakai-Sugimoto model. We calculate the classical instanton solution and use collective coordinate method to obtain the wave function of baryons. In this model, baryons appear as quantized five-dimensional instanton. 

The Sakai-Sugimoto model is an effective model of hadrons based on gauge/string duality. It is described by the $U(N_f)$ Yang--Mills--Chern--Simons theory in curved five-dimensional space-time, where $N_f$ is the number of flavors. The action is given by the Yang--Mills part $S_{\mathrm{YM}}$ and Chern--Simons part $S_{\mathrm{CS}}$ as follows:
	\begin{align} 
S=&S_{\mathrm{YM}}+S_{\mathrm{CS}}, \label{eq:01}\\
S_{\mathrm{YM}}=&-\kappa \int d^{4} x d z \operatorname{tr}\left[\frac{1}{2} h(z) \mathcal{F}_{\mu \nu}^{2}+k(z) \mathcal{F}_{\mu z}^{2}\right], \\
S_{\mathrm{CS}}=&\frac{N_{c}}{24 \pi^{2}} \int_{M^{4} \times \mathbb{R}} \omega_{5}(\mathcal{A}).
	\end{align}
Here, $\mu,\nu=0,1,2,3$ are four-dimensional Lorentz indices and $z$ is the coordinate of the fifth dimension. The warp factors $k(z)$ and $h(z)$ are defined as $k(z)=1+z^2$ and $h(z)=(1+z^2)^{-1/3}$, respectively. The number of colors $N_c$ is related to $\kappa$ as
	\begin{equation*}
		\kappa = \frac{\lambda N_c}{216\pi^3}
	\end{equation*}
where $\lambda$ is the 't Hooft coupling. $\mathcal{A}=\mathcal{A}_\alpha dx^\alpha = \mathcal{A}_\mu dx^\mu + \mathcal{A}_z dz\ (\alpha = 0,1,2,3,z)$ is the five-dimensional $U(N_f)$ gauge field and $\mathcal{F}=\frac{1}{2}\mathcal{F}_{\alpha\beta} dx^\alpha \wedge dx^\beta = d\mathcal{A} + i\mathcal{A}\wedge\mathcal{A}$ is the field strength. In this paper, we consider the case of $N_f = 2$. We decompose $U(2)$ gauge field as 
\begin{equation*}
		\mathcal{A} = A^a\frac{\tau^a}{2} + \hat{A}\frac{I_2}{2}, 
	\end{equation*}
where $\hat{A}$ is the $U(1)$ part, $A$ the $SU(2)$ part, $\tau^a\ (a=1,2,3)$ the Pauli matrices and $I_2$ the 2$\times$2 unit matrix. The Chern--Simons five-form is
	\begin{equation*}
		\omega_{5}(\mathcal{A})=\operatorname{tr}\left(\mathcal{A} \mathcal{F}^{2}-\frac{i}{2} \mathcal{A}^{3} \mathcal{F}-\frac{1}{10} \mathcal{A}^{5}\right).
	\end{equation*}
In this paper, we mainly work with the unit of the Kaluza--Klein mass $M_{\mathrm{KK}} = 1$.

Because of the warp factors $k(z)$ and $h(z)$, it is difficult to solve the equation of motion in general. However, the size is small in the large $\lambda$ limit, since the instanton size is proportional to $\lambda^{-1/2}$. Then, we can consider the solution localized at the origin $z\sim 0$. In this case, we can set $k(z)=h(z)=1$ and the solution is given by the BPST instanton solution:
	\begin{align*}
&A_{M}^{c l}(\bm{x}, z)=-i f(\xi) g \partial_{M} g^{-1}, \\
&A_{0}^{c l}=0, \\
&\hat{A}_{M}^{c l}=0, \\
&\hat{A}_{0}^{c l}=\frac{1}{8 \pi^{2} a} \frac{1}{\xi^{2}}\left[1-\frac{\rho^{4}}{\left(\xi^{2}+\rho^{2}\right)^{2}}\right]
	\end{align*}
with 
	\begin{align*}
		&f(\xi)=\frac{\xi^{2}}{\xi^{2}+\rho^{2}}, \quad g(x)=\frac{(z-Z)-i(\bm{x}-\bm{X}) \cdot \bm{\tau}}{\xi},\\
		 &\xi=\sqrt{(z-Z)^{2}+|\bm{x}-\bm{X}|^{2}}
	\end{align*}
where $X^M=(X^1,X^2,X^3,Z)=(\bm{X},Z)$ is the positions of the instanton in four-dimensional space and $\rho$ is a parameter of the instanton size.

Let us quantize the classical instanton solution by the collective coordinate method. In order to do that, we introduce  time--dependent parameters $\bm{X}(t),Z(t),\rho (t)$ and $SU(2)$ orientation $V(t,x,a(t))$ with $V(z\to \pm \infty) \to \bm{a}(t)$, where $\bm{a}(t)=a_4(t) + ia_a(t)\tau^a$ is a variable representing rotations in spin or isospin space. In this case, the $SU(2)$ gauge field is given by 
	\begin{equation*}
		A_M(t,x)=VA_M^{\mathrm{cl}}(x;X(t),\rho(t))V^{-1} - iV\partial_MV^{-1}
	\end{equation*}
where $A_M^{\mathrm{cl}}$ is the classical solution and $V$ satisfies
	\begin{align*}
		-i V^{-1} \dot{V}&=-\dot{X}^{M}(t) A_{M}^{\mathrm{cl}}+\chi^{a} f(\xi) g \frac{\tau^{a}}{2} g^{-1},\\
		\chi^{a}&=-i \operatorname{tr}\left(\tau^{a} \boldsymbol{a}^{-1} \dot{\boldsymbol{a}}\right).
	\end{align*}
We insert this gauge field in (\ref{eq:01}) and integrate with respect to $(x^\mu,z)$. As a result, the instanton is quantized by the collective coordinate method. We obtain the collective hamiltonian 
	\begin{align*}
		&H=-\frac{1}{2 M_{0}}\left(\partial_{\bm{X}}^{2}+\partial_{Z}^{2}\right)-\frac{1}{4 M_{0}} \partial_{y^{I}}^{2}+U(\rho, Z), \\
		&U(\rho, Z)=M_{0}+\frac{M_{0}}{6} \rho^{2}+\frac{N_{c}^{2}}{5 M_{0}} \frac{1}{\rho^{2}}+\frac{M_{0}}{3} Z^{2},
	\end{align*}
where $M_0 = 8\pi^2\kappa$ is the instanton mass and $y_I = \rho a_I\ (I=0,1,2,3)$.

The eigenstates of this Hamiltonian are characterized by quantum numbers $(l,I_3,s_3,n_\rho,n_z)$ and momentum $\bm{p}$: $l=1,3,5,\cdots$ are related to spin $J$ and isospin $I$ as $l/2=J=I$. $I_3$ and $s_3$ are third components of the isospin and spin, respectively, and $n_\rho$ and $n_z$ the quantum numbers representing the oscillations in the radial and $z$ directions, respectively. The ground state with $n_\rho=n_z=0$  is the nucleon. In addition, the first excited states of $n_\rho=1$ or of $n_z=1$ corresponds to the $N(1440)$ or $N(1535)$, respectively. The wave functions of the ground states and the $N(1535)$ resonant state of the spin-upward proton are as follows:
	\begin{align}
		\begin{split}
			\psi_{N} & \propto e^{i \bm{p} \cdot \bm{X}} R_{N}(\rho) \psi_{Z}(Z)\left(a_{1}+i a_{2}\right) ,\\
			\psi_{N^*} & \propto e^{i \bm{p} \cdot \bm{X}} R_{N}(\rho) \psi'_{Z}(Z)\left(a_{1}+i a_{2}\right) \label{eq:psi}
		\end{split}
	\end{align}
where $N^*=N(1535)$ and
\begin{align*}
	&R_{N}(\rho)=\rho^{-1+2 \sqrt{1+N_{c}^{2} / 5}} e^{-\frac{M_{0}}{\sqrt{6}} \rho^{2}} ,\\
	&\psi_{Z}(Z)=e^{-\frac{M_{0}}{\sqrt{6}} Z^{2}},\\
	&\psi'_Z(Z) = Z\psi_Z(Z).
\end{align*}

\subsection{The asymptotic solution of the instanton}
In order to calculate the coupling constant of the interaction, we need to calculate the current. To define the current in the holographic model, we need to investigate the physics of the boundary based on the Gubser--Klebanov--Polyakov--Witten (GKP-W) method~\cite{GKP:1998, Witten:1998}.

We have already seen the BPST instanton solution which is localized at $z \sim 0$ in the large $\lambda$ limit. However, as we will see later, the current is defined at $z \to \pm \infty$, and therefore this solution is not appropriate. We need to investigate properly the behavior of the soliton solution in a large $z$ region. It is given in Ref.~\cite{Hashimoto:2008zw} as follows:
\begin{widetext}
	\begin{align*}
		\hat{A}_{0}=&-\frac{1}{2 a \lambda} G(\bm{x}, z ; \bm{X}, Z), \\
		\hat{A}_{i}=&\frac{1}{2 a \lambda}\left[\dot{X}^{i}+\frac{\rho^{2}}{2}\left(\frac{\chi^{a}}{2}\left(\epsilon^{i a j} \frac{\partial}{\partial X^{j}}-\delta^{i a} \frac{\partial}{\partial Z}\right) \right.
		\left. +\frac{\dot{\rho}}{\rho} \frac{\partial}{\partial X^{i}}\right)\right] G(\bm{x}, z ; \bm{X}, Z), \\
		\hat{A}_{z}=&\frac{1}{2 a \lambda}\left[\dot{Z}+\frac{\rho^{2}}{2}\left(\frac{\chi^{a}}{2} \frac{\partial}{\partial X^{a}}+\frac{\dot{\rho}}{\rho} \frac{\partial}{\partial Z}\right)\right] H(\bm{x}, z ; \bm{X}, Z), \\
		A_{0}=&4 \pi^{2} \rho^{2} i a \dot{a}^{-1} G(\bm{x}, z ; \bm{X}, Z) +2 \pi^{2} \rho^{2} a \tau^{a} a^{-1}\left(\dot{X}^{i}\left(\epsilon^{i a j} \frac{\partial}{\partial X^{j}} \right.\right. 
		\left.\left. -\delta^{i a} \frac{\partial}{\partial Z}\right)+\dot{Z} \frac{\partial}{\partial X^{a}}\right) G(\bm{x}, z ; \bm{X}, Z), \\
		A_{i}=&-2 \pi^{2} \rho^{2} \boldsymbol{a} \tau^{a} \boldsymbol{a}^{-1}\left(\epsilon^{i a j} \frac{\partial}{\partial X^{j}}-\delta^{i a} \frac{\partial}{\partial Z}\right) G(\bm{x}, z ; \bm{X}, Z) ,\\
		A_{z}=&-2 \pi^{2} \rho^{2} \boldsymbol{a} \tau^{a} \boldsymbol{a}^{-1} \frac{\partial}{\partial X^{a}} H(\bm{x}, z ; \bm{X}, Z) .
	\end{align*}
\end{widetext}
Here, $i=1,2,3$, $\epsilon^{abc}$ is the completely antisymmetric tensor and $\delta^{ab}$ the Kronecker delta  . Green's functions $G(\bm{x}, z ; \bm{X}, Z), H(\bm{x}, z ; \bm{X}, Z)$ are
	\begin{align*}
		&G(\bm{x}, z ; \bm{X}, Z)=\kappa \sum_{n=1}^{\infty} \psi_{n}(z) \psi_{n}(Z) Y_{n}(|\bm{x}-\bm{X}|), \\
		&H(\bm{x}, z ; \bm{X}, Z)=\kappa \sum_{n=0}^{\infty} \phi_{n}(z) \phi_{n}(Z) Y_{n}(|\bm{x}-\bm{X}|)
	\end{align*}
with 
	\begin{align*}
		&\phi_{0}(z)=\frac{1}{\sqrt{\kappa \pi}} \frac{1}{k(z)}, \\
		&\phi_{n}(z)=\frac{1}{\sqrt{\lambda_{n}}} \partial_{z} \psi_{n}(z), \\
		&Y_{n}(r)=-\frac{1}{4 \pi} \frac{e^{-\sqrt{\lambda_{n}}} r}{r}, \quad r=|\bm{x}|.
	\end{align*}
Here $\{\psi_n(z)\}\quad(n=1,2,\dots)$ is a complete set of functions of $z$ consisting of the eigenfunctions of the eigenequation
	\begin{equation}
		-h(z)^{-1} \partial_z (k(z) \partial_z \psi_n) = \lambda_n \psi_n \label{eq:difeq}
	\end{equation}
  with the normalization condition
	\begin{equation*}
		\kappa\int dz h(z)\psi_n\psi_m = \delta_{mn}.
	\end{equation*}
We note that $\psi_{2n-1}(Z)$ and $\psi_{2n}(Z)$ are even and odd function of $Z$, respectively.

\subsection{Currents}
In Ref ~\cite{Hashimoto:2008zw} , the left and right currents $\mathcal{J}^\mu_L, \mathcal{J}^\mu_R$ are defined by coupling with external fields $\mathcal{A}_{L\mu}, \mathcal{A}_{R\mu}$ respectively at $z=\pm \infty$: 
	\begin{equation*}
		\left.S\right|_{\mathcal{O}\left(\mathcal{A}_{L}, \mathcal{A}_{R}\right)}=-2 \int d^{4} x \operatorname{tr}\left(\mathcal{A}_{L \mu} \mathcal{J}_{L}^{\mu}+\mathcal{A}_{R \mu} \mathcal{J}_{R}^{\mu}\right).
	\end{equation*}
We introduce the external fields in the GKP--W method:
\begin{align*}
		&\mathcal{A}_{\alpha}\left(x^{\mu}, z\right)=\mathcal{A}_{\alpha}^{c l}\left(x^{\mu}, z\right)+\delta \mathcal{A}_{\alpha}\left(x^{\mu}, z\right),\\
		&\delta \mathcal{A}_{\mu}\left(x^{\nu}, z \rightarrow+\infty\right) =\mathcal{A}_{L \mu}\left(x^{\nu}\right) ,\\
		&\delta \mathcal{A}_{\mu}\left(x^{\nu}, z \rightarrow-\infty\right) =\mathcal{A}_{R \mu}\left(x^{\nu}\right)
	\end{align*}
where $\mathcal{A}^{\mathrm{cl}}$ is the classical solution and $\delta\mathcal{A}_\alpha$ an infinitesimal deviation. Substituting this fields in the action and taking only liner terms, we obtain
	\begin{equation*}
		\left.S\right|_{\mathcal{O}\left(\mathcal{A}_{L}, \mathcal{A}_{R}\right)}=\kappa \int d^{4} x 2 \operatorname{tr}\left[\operatorname{tr}\left(\delta \mathcal{A}^{\mu} k(z) \mathcal{F}_{\mu z}^{\mathrm{cl}}\right)\right]_{z=-\infty}^{z=+\infty},
	\end{equation*}
and so
	\begin{align*}
		\mathcal{J}_{L \mu}=& -\left.\kappa\left(k(z) \mathcal{F}_{\mu z}^{\mathrm{cl}}\right)\right|_{z=+\infty}, \\\quad \mathcal{J}_{R \mu}=& +\left.\kappa\left(k(z) \mathcal{F}_{\mu z}^{\mathrm{cl}}\right)\right|_{z=-\infty}.
	\end{align*}
The axial current is defined by 
	\begin{equation*}
		\mathcal{J}_{A}^{\mu}=\mathcal{J}_{L}^{\mu}-\mathcal{J}_{R}^{\mu}=-\kappa\left[\psi_{0}(z) k(z) \mathcal{F}_{\mu z}^{\mathrm{cl}}\right]_{z=-\infty}^{z=+\infty},
	\end{equation*}
with $\psi_0(z) = (2/\pi)\arctan z$.

We define 
	\begin{equation*}
		\mathcal{J}_A^\mu =  J_A^{a\mu} \frac{\tau^a}{2} + \hat{J}_A^\mu \frac{I_2}{2}
	\end{equation*}
and use the asymptotic solutions. Thus we obtain
	\begin{align*}
		J_{A}^{0}&(r ; \bm{X}, Z, \rho, \bm{a})\\
		&= 2 \pi^{2} \kappa\left[\partial_{0}\left(\rho^{2} \boldsymbol{a} \tau^{a} \boldsymbol{a}^{-1}\right) \partial_{a} H^{A}-2 \rho^{2} i \boldsymbol{a} \dot{\boldsymbol{a}}^{-1} G^{A}\right.\\
		&\left.-\rho^{2} \boldsymbol{a} \tau^{a} \boldsymbol{a}^{-1} \dot{X}^{i}\left(\left(\partial_{a} \partial_{i}-\delta^{i a} \partial_{j}^{2}\right) H^{A}-\epsilon^{i a j} \partial_{j} G^{A}\right)\right],\\
		J_{A}^{i}&(r ; \bm{X}, Z, \rho, \bm{a})\\
		&=-2 \pi^{2} \kappa \rho^{2} \boldsymbol{a} \tau^{a} \boldsymbol{a}^{-1} \left(\left(\partial_{i} \partial_{a}-\delta^{i a} \partial_{j}^{2}\right) H^{A}-\epsilon^{i a j} \partial_{j} G^{A}\right)
	\end{align*}
where
	\begin{align*}
		G^{A}(Z, r) \equiv& \left[\psi_{0}(z) k(z) \partial_{z} G\right]_{z=-\infty}^{z=+\infty}\\
		=&-\sum_{n=1}^{\infty} g_{a^{n}} \psi_{2 n}(Z) Y_{2 n}(r),\\
		H^{A}(Z, r) \equiv& \left[\psi_{0}(z) k(z) H\right]_{z=-\infty}^{z=+\infty}\\
		=&-\frac{1}{2 \pi^{2}} \frac{1}{k(Z)} \frac{1}{r}-\sum_{n=1}^{\infty} \frac{g_{a^{n}}}{\lambda_{2 n}} \partial_{Z} \psi_{2 n}(Z) Y_{2 n}(r),\\
		g_{a^{n}}=&\lambda_{2 n} \kappa \int d z h(z) \psi_{2 n} \psi_{0}
	\end{align*}
and $r = |\bm{x} - \bm{X}|$. Note that $G^A$ and $H^A$ are even and odd function of $Z$, respectively.
In momentum space, we have the axial current as follows:
\begin{widetext}
	\begin{align}
		\tilde{J}_{A}^{\mu}(\bm{k})=&\int d^{3} x e^{-i \bm{k} \cdot \bm{x}} J_{A}^{\mu}(r), \\
		\tilde{J}_A^{a0}(\bm{k}) =& 2\pi^2\kappa \left\{
			i\frac{k_a}{\bm{k}^2} \mathrm{tr} [\tau^a \partial_0 (\rho^2 \bm{a} \tau^b \bm{a}^{-1})] \right.
			\left. + \frac{P_i}{M_0}\rho^2 \mathrm{tr}[\tau^a\bm{a}\tau^b\bm{a}^{-1}]\left( \delta^{ib} - \frac{k_ik_b}{\bm{k}^2}\right)
			\right\}e^{-ikX} \sum_{n=1}^{\infty} \frac{g_{a^n}\partial_Z\psi_{2n}}{\bm{k}^2 + \lambda_{2n}} \nonumber \\
			&+ \left( I_a - 2\pi^2\kappa\rho^2\frac{P_i}{M_0}\epsilon^{iaj}k_j \right) e^{-ikX}\sum_{n=1}^{\infty}\frac{g_{a^n}\psi_{2n}}{k^2+\lambda_{2n}}, \\
			\widetilde{J}_{A}^{a i}(\bm{k})=& 2 \pi^{2} \kappa \rho^{2} \operatorname{tr} \left(\tau^{a} \boldsymbol{a} \tau^{b} \boldsymbol{a}^{-1}\right) e^{-i \bm{k} \cdot \bm{X}}
			\left[\left(\delta_{b i}-\frac{k_{b} k_{i}}{\bm{k}^{2}}\right) \sum_{n = 1}^\infty \frac{g_{a^{n}} \partial_{Z} \psi_{2 n}}{\bm{k}^{2}+\lambda_{2 n}}
			-i\epsilon^{ibj}k_j\sum_{n=1}^\infty \frac{g_{a^n}\psi_{2n}}{\bm{k}^2+\lambda_{2n}} \right], \label{eq.spacial}
	\end{align}
\end{widetext}
where
	\begin{align*}
		&P_i=M_0 \dot{X}_i ,\\
		&I_{a}=\frac{i}{2}\left(y_{4} \frac{\partial}{\partial y_{a}}-y_{a} \frac{\partial}{\partial y_{4}}-\epsilon_{a b c} y_{b} \frac{\partial}{\partial y_{c}}\right).
	\end{align*}

\section{DECAY PROPERTIES OF N(1535)}
\subsection{Axial coupling}
Let us calculate the axial coupling of the $N(1535)\to N\pi$ decay process using the current we have calculated. It is given by
\begin{equation}
g_A^{NN^*}(\tau^a)_{I_3I'_3} = 2\int d^3x \braket{N,I'_3|J_A^{a0}|N^*,I_3}e^{i\bm{k}\cdot\bm{x}} \label{eq. g_A}.
\end{equation}
As we can see from (\ref{eq.spacial}), the spatial components of the currents $\widetilde{J}_{A}^{a i}(\bm{k})$ do not contribute to the axial coupling because of the parity of the wave functions $\psi(z), \psi'(z)$ in (\ref{eq:psi}) with respect to $Z$ and completely antisymmetric tensor. This fact is consistent with the following property. In the non-relativistic limit, the axial current between $N$ and $N^*$ has the structure
\begin{align*}
	J^\mu_A=\bar{\psi}_N \gamma^\mu \psi_{N^*}, \quad \psi_B  = \begin{pmatrix} u_B\\0 \end{pmatrix},\quad B=N,N^*.
\end{align*}
Note that $\gamma_5$ does not appear because of the difference in the paritys of the initial and final states. Then we see that spatial component vanishes while the time component contributes.

The axial coupling constant define at $\bm{k}=0$. Calculating (\ref{eq. g_A}), we obtain
\begin{align*}
		g_A^{NN^*} =& \int_{-\infty}^\infty dZ \psi_Z(Z)\psi'_Z(Z) \sum_{n=1}^\infty \frac{g_{a^n} \psi_{2n}(Z)}{\lambda_{2n}} 
\end{align*}
where $\psi_Z(Z)$ and $\psi'_Z(Z)$ are normalized wave functions.
Using
	\begin{equation}
		\sum_{n=1}^\infty \frac{g_{a^n} \psi_{2n}(Z)}{\lambda_{2n}} = \frac{2}{\pi}\arctan Z = \psi_0(Z) ,\label{eq:psi0}
	\end{equation}
we get
\begin{equation*}
		g_A^{NN^*} = \int_{-\infty}^\infty dZ \psi_Z(Z) \psi'_Z(Z) \psi_0(Z).
\end{equation*}
We obtain analytically with the result
	\begin{equation}
		g_A^{NN^*} = \sqrt{\frac{
		2}{\pi}} e^{\frac{2}{\sqrt{6}}M_0}
		\mathrm{erfc} \left( \sqrt{\frac{
		2}{\sqrt{6}}M_0} \right) \label{eq:gA}
	\end{equation}
where $\mathrm{erfc(x)}$ is the complementary error function.

For numerical estimate, we choose the parameters
	\begin{align*}
		\kappa =& 0.00745,\\
		M_{KK} =& 940\ \mathrm{MeV},
	\end{align*}
which are related to 
	\begin{align*}
		M_0 =& 8\pi^2\kappa = 0.588, \\
		\kappa =& \frac{\pi}{4}\frac{f_\pi^2}{M_{KK}^2}, 
	\end{align*}
and so $f_\pi = 92.4\ \mathrm{MeV}$.
$M_{KK}$ is determined to reproduce the mass of $\rho \ (776\ \mathrm{MeV})$ ~\cite{Sakai:2004cn}. 
Using the parameters, we obtain the value 
	\begin{align}
		g_A^{NN^*} = 0.42.
		\label{gANN*}
	\end{align}

\subsection{Decay width}
Let us calculate the decay width of the $N(1535)\to N\pi$ decay proses from the axial coupling we calculated, and compare it with the experimental value.

The decay width is given by 
	\begin{align*}
		\Gamma_{N^*\to N\pi} = \frac{1}{2 m_{N^*}}&\int \frac{d^{3} p_{N}}{(2 \pi)^{3} 2 E_{N}} \frac{d^{3} k}{(2 \pi)^{3} 2 E_{\pi}} \notag\\
		&\times(2 \pi)^{4} \delta^{4}\left(p_{N}+k-p_{N^*}\right)\left|t_{f i}\right|^{2}, \label{eq:Gamma}
	\end{align*}
with
\begin{align*}
		t_{fi} =& \braket{N(-k); \pi(k)|\mathcal{L}|N^*(0)}\\
		=& i \frac{g_A^{NN^*}}{2 f_\pi} E_\pi \sqrt{E_N + m_N} \sqrt{2m_{N^*}} \delta_{s_3s'_3},\\
		\mathcal{L} =& i \frac{g_A^{NN^*}}{2 f_\pi} \bar{\psi}_N \gamma^0 \partial_0 \pi^a \tau^a \psi_{N^*},
	\end{align*}
where $m_{N^*}$ is the mass of $N(1535)$, $m_N$ the nucleon mass, $E_N$ the energy of nucleon, $E_\pi=\sqrt{\bm{k}^2 + m^2_\pi}$ and $m_\pi$  the pion mass. Hence, we obtain 
	\begin{equation*}
		\Gamma_{N^*\to N\pi} = \frac{g_A^{NN^*2}}{16\pi} \frac{|\bm{k}|E_\pi^2(E_N + m_N)}{f_\pi^2 m_{N^*}},
	\end{equation*}
	where $\bm{k}$ is the momentum of the emitted pion
	\begin{align*}
		|\bm{k}|=&\left[\frac{m_{N^*}^4 + m_{N}^4 + m_{\pi}^4}{4m_{N^*}^2}\right.\\
		&\left.-\frac{ 2(m_{N^*}^2 m_{N}^2 + m_{N}^2 m_{\pi}^2 + m_{\pi}^2 m_{N^*}^2)}{4 m_{N^*}^2}\right]^{1/2}.
	\end{align*}
Here, to obtain $\pi N N^*$ coupling constant, we have used the Goldberger--Treiman relation
	\begin{align*}
		g^{\pi NN^*} = \frac{m_{N^*}-m_N}{2f_\pi}g_A^{NN^*}.
	\end{align*}

From PDG, $m_{N^*}=1510\ \mathrm{MeV}$, $m_N=940\ \mathrm{MeV}$, $m_\pi = 140\ \mathrm{MeV}$, and $|\bm{k}| =448\ \mathrm{MeV}$, where the mass of $N(1535)$ is the pole position. 
Using these kinematical values and $g_A^{NN^*}$ as given in (\ref{gANN*})
we obtain the decay width
	\begin{equation*}
		\Gamma_{N^*\to N\pi} = 54\ \mathrm{MeV}. \label{eq:width}
	\end{equation*}
The experimental value of the PGD is that the total decay width is 130 MeV and the branching rate is 32 -- 52 \%, then we obtain
	\begin{equation*}
		\Gamma_{N^*\to N\pi}^\mathrm{exp} = 42 - 68\ (\approx 55)\ \mathrm{MeV} 
	\end{equation*}
which is in good agreed. 

\begin{figure}[hbtp]
	\includegraphics[scale=0.5]{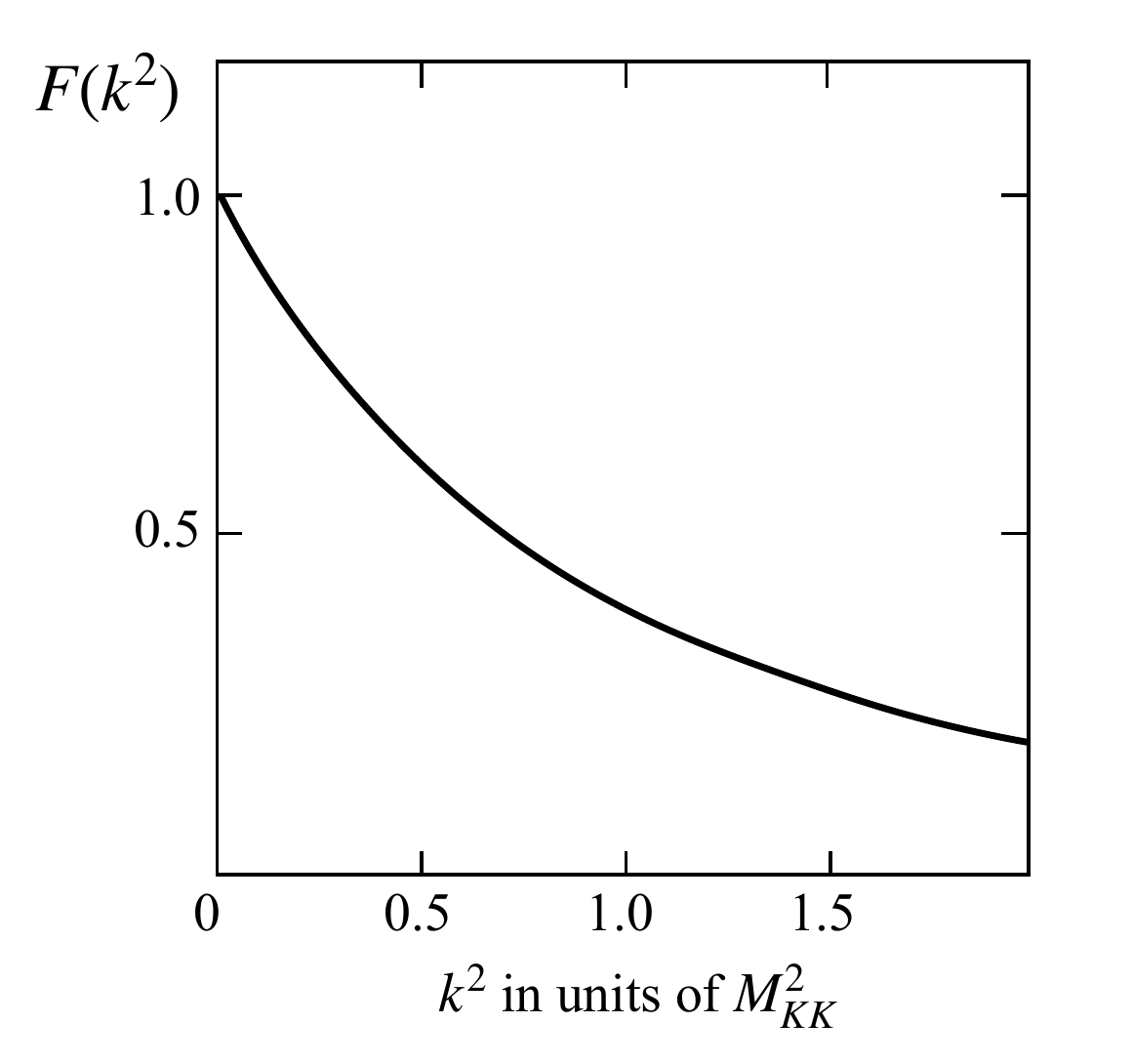}
	\caption{The emitted pion momentum $\bm{k}$ dependance of the form factor $F_A(\bm{k}^2)$}
	\label{fig.1}
\end{figure}

In the above calculations we have used the value 
of $g_A^{NN^*}$ at the zero momentum $\bm{k}$ of the emitted pion. 
To compute the decay width at a finite $\bm{k}$, we need to 
take into account the form factor.  
The form factor is given by
\begin{align*}
	\langle F_A(Z,\bm{k}^2) \rangle = \frac{1}{g_A^{NN^*}} \sum_{n=1}^\infty \frac{g_{a^n} \langle \psi_{2n}(Z) \rangle}{\bm{k}^2 + \lambda_{2n}}
\end{align*}
where the bracket symbol $\langle \rangle$ represents the matrix element with the wave functions of the nucleon and $N(1535)$ states (\ref{eq:psi}).  
We have numerically computed the sum over $n$ and verified a good convergence for the maximum value 
$n \sim 15$.  
The result is shown in Fig.\ref{fig.1} as a function of $k^2 = |\bm{k}|^2$ in units of $M_{KK}^2$.  
From this figure, we can read the actually emitted pion momentum $|\bm{k}| = 448\ \mathrm{MeV}$ corresponds to $k^2 \sim 0.25$, where $F(k^2) \sim 0.75$.  
Thus, the value of the axial coupling $g_A^{NN^*}$ at this pion momentum is about $g_A^{NN^*}(|\bm{k}|\sim 450\ \mathrm{MeV})\sim 0.32$ and the resulting decay width is $\Gamma_{N^*\to N\pi}(|\bm{k}|\sim 450\ \mathrm{MeV}) \sim 30\ \mathrm{MeV}$.
This value is slightly smaller than the experimental value, which however is still acceptable.   

By using the form factor we can also estimate the distribution (size) of the axial density, the axial radius.  
It is defined by 
\begin{equation*}
	\frac{\partial F_A(\bm{k}^2)}{\partial \bm{k}^2} = - \frac{1}{6} \langle r^2 \rangle_A
\end{equation*}
which is estimated to be $\langle r^2 \rangle^{1/2} \sim 0.5$ fm.
This corresponds to the axial mass around 1 GeV $\sim \sqrt{6} \langle r^2 \rangle_A^{-1/2}$.
This value is close ot the mass of the axial vector meson which is expected to be 
$m_{a_1} = \sqrt{2} m_\rho \sim 1080$ MeV~\cite{Weinberg:1969hw}.

\section{SUMMARY AND DISCUSSIONS}
In this paper, the one-pion emission decay process $N(1535) \to \pi N$ is investigated using the Sakai-Sugimoto model, which is a model of holographic QCD. In this model, baryons are formed as instantons in 5-dimensional spacetime. We defined the axial current according to the formulation of ~\cite{Hashimoto:2008zw} and calculated the value of the axial coupling for the $N(1535) \to \pi N$ process. Using the obtained axial coupling, we calculated the decay width of the $N(1535) \to \pi N$ process and obtained a value of 54 MeV. This value is in good agreement with the experimental value. 

In general, when the axial coupling constants $g_A$ of various baryons are calculated using a soliton picture-based model, the obtained values are known to be about 30 \% smaller ~\cite{Adkins:1983ya}. This tendency is also observed in the present and the recent studies for resonance properties.  
In Ref.~\cite{Fujii:2021}, the decay of the Roper resonance $N(1440) \to \pi N$ was estimated to be 64 MeV
which is some smaller than the corresponding experimental partial decay width, 90-140 MeV.  
Nevertheless, it is amusing that the holographic approach explains qualitatively well dynamical properties of 
baryon resonances as well as static properties.  

\begin{acknowledgments}
This work was supported in part by Grants-in-Aid for Scientific Research [No. 21H04478(A)] 
and the one on Innovative Areas (No. 18H05407).
\end{acknowledgments}

\appendix

\bibliography{N1535}

\end{document}